# The Problems of Personnel Training for STEM Education in the Modern Innovative Learning and Research Environment

Mariya P. Shyshkina[0000-0001-5569-2700]

Institute of Information Technologies and Learning Tools of NAES of Ukraine,
9, M. Berlynskoho St., Kyiv, 04060, Ukraine
shyshkina@iitlt.gov.ua

**Abstract.** The *aim of the article* is to describe the problems of personnel training that arise in view of extension of the STEM approach to education, development of innovative technologies, in particular, virtualization, augmented reality, the use of ICT outsourcing in educational systems design. The *object of research* is the process of formation and development of the educational and scientific environment of educational institution. The *subject of the study* is the formation and development of the cloud-based learning and research environment for STEM education. The *methods of research* are: the analysis of publications on the problem; generalization of domestic and foreign experience; theoretical analysis, system analysis, systematization and generalization of research facts and laws for the development and design of the model of the cloud-based learning environment, substantiation of the main conclusions. The *results of the research* are the next: the concepts and the model of the cloud-based environment of STEM education is substantiated, the problems of personnel training at the present stage are outlined.

## 1 Introduction

Nowadays it is impossible to introduce the innovative ICT into the learning process and management of pedagogical systems without paying attention to the organization of teachers training in educational institutions especially in view of the current need of a large amount of highly skilled IT personnel for information society development. To train teachers that would be involved in the process of informatization of the modern educational environment it is necessary to develop new approaches related to the education at different level and profile of training [7].

There is a significant need in ICT-qualified specialists in the field of public administration of education, educational management, training and retraining of teaching staff. Without sufficient knowledge of the current developments of educational ICT services the graduates will have problems with adaptation at their workplace due to the lack of awareness of the real issues and working conditions of innovative ICT use, as well as the lack of ideas on the practical implementation of innovations in the educational process, low level of emerging educational techniques introduction [6; 7].

It is unlikely that the present state of scientific, educational and management personnel training is quite satisfactory for the needs of innovative development of ICT for learning both regarding the required number of qualified specialists and the content and quality of training. Therefore there is a need for the development of new models and approaches to personnel training in view of the modernization of ICT infrastructure and the integration of learning resources at different levels of education, management and research [6; 7].

## 2  Results and Discussion

The process of creation and content elaboration of electronic learning resources requires fundamental basic knowledge in the field of computer and educational technologies. Instead, approaches to training today are usually not enough focused on innovations that has taken place in recent years and on the real needs for such training [7]. Certain approach to these problems solving can be based on the mechanism of outsourcing of ICT services provision with the use of appropriate cloud computing services [1]. Outsourcing plays a significant role in improving the technical level of ICT systems of educational institutions, as well as the efficiency of their processing and development. It is a market mechanism enforcing the introduction of the latest advances in ICT, aimed at more flexible and prompt response to the needs of the user [1]. In view of the challenges of educational institution innovative ICT infrastructure formation it would be possible to solve some of the above-mentioned problems [6].

Under *the learning and research environment* of the educational institution the environment of the learning and research activities of its participants (students, listeners, teachers, methodologists, scientists, administrative, managerial and auxiliary staff) is meant, where the necessary, sufficient and safe conditions for its implementation are created.

*The cloud-based learning and research environment* of educational institution is the environment of the learning and research activities of the participants, where the virtualized computer-technological (corporate or hybrid) infrastructure is purposefully created to provide its computer-processing functions.

The cloud-based approaches to the formation of the educational environment have promising application in the field of STEM education (Science, Technology, Engineering, Mathematics). The research in this area is aimed at achieving of a new quality of learning by means of more powerful, flexible, scalable infrastructure solutions that can be used to integrate a variety of educational components based on emerging technologies into the learning and research environment. Thus, the concept of the cloud-based environment of STEM-education appears to be valuable.

Among the functions of the cloud-based environment are: support for various processes of learning and research activities within an educational institution, supply of educational resources and services based on a unite platform.

This leads to the concept of cloud-based learning technology namely the computer-oriented part of learning technology aimed at solving of certain didactic tasks. It reflects the model of learning structure (as a set of relationships among the participants of the learning process, the elements of content and other components of the computer-based environment) the use of computer-based learning tools, information and communication networks and electronic resources, mainly and fundamentally based on cloud computing [7].

Recently in the field of STEM education the following ICT trends [4] have been developed, such as new interfaces, screenless displays, 3D technologies, augmented reality [2], "emotional" computing, wearable technologies (devices) and others. All these areas are united under the common name of "new opportunities" (new enabling technologies) [3; 5].

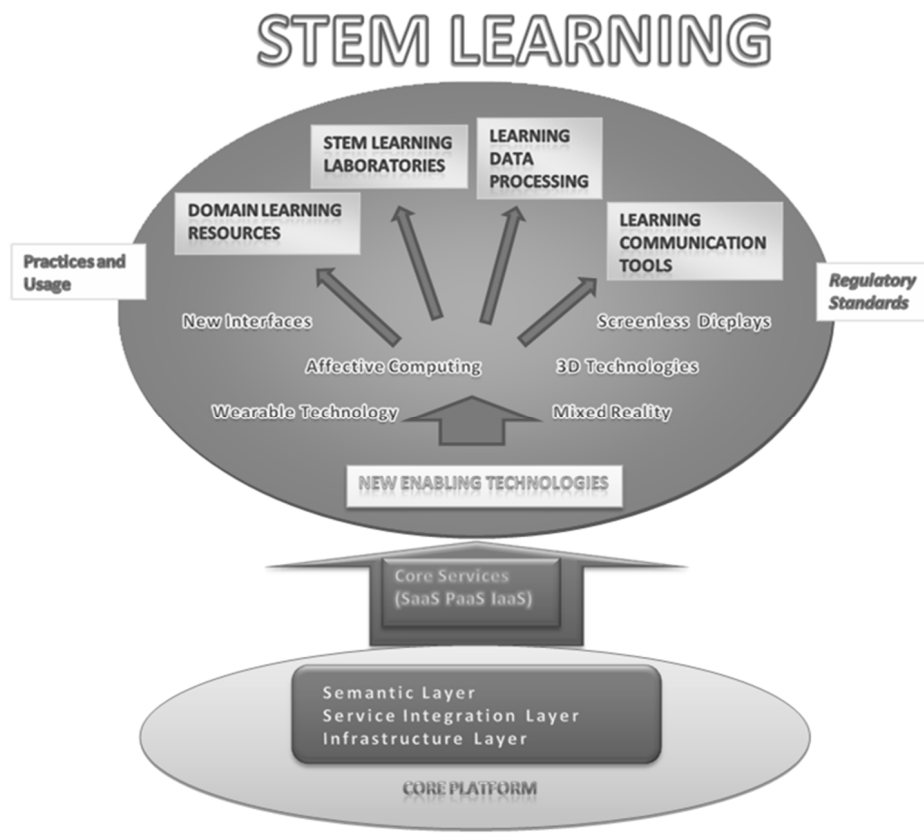

**Fig. 1.** The model of the cloud-based learning and research environment of STEM-education

If we consider STEM-education tools and services in terms of the cloud-based learning and research environment services types, it is possible to distinguish the following elements that are its most important structural units:

- the personalized (remote) STEM education laboratories, which contains the tools for the management of various specialized software, devices and equipment through the network;
- the subject-oriented collections, libraries, learning resources depositories, containing sets of different programs and data for educational purposes;
- the specialized corporate cloud software, in particular, those services of modeling, programming, computing, designing, solving of educational tasks that are available for a corporate range of users – for example, employees and students of an educational institution;
- the services of scientific and educational information networks that can provide access to various data, electronic resources and network tools for scientific and educational purposes, provided to the participants of a public network (Fig. 1).

Among the computer-based tools for creating, combining and reusing of content, services, applications and data in STEM learning process there are such as simulation tools; embedded network objects; platforms and networks for the organization of joint activities; communication tools in the learning process and others. In [1] the following components are distinguished:

- the environment for learning objects testing and experimentation ("residence" learning experience), for example, using 3D-modeling, imaging technology, augmented and virtual reality, adaptive / personalized environments);
- the learning support services (e.g., data processing, training analytics for tracking and evaluating dynamically the real-time student's achievements) [3].

## 3    Conclusion

Thus, the problems of STEM-education personnel training in modern innovative environment are in much concern along with the need for wider implementation of ICT-outsourcing in the design of educational systems and the development deployment of innovative technologies, including virtualization, augmented reality and others in the processes of teachers training.